\documentclass[11pt]{amsart}
\usepackage[english]{babel}
\usepackage{xcolor}
\usepackage{pgf}
\usepackage{geometry}
\usepackage{times}
\usepackage{graphicx,psfrag,etoolbox}
\usepackage[numbers]{natbib}
\usepackage{csquotes}
\usepackage{dsfont}

\newtheorem{thm}{Theorem}[section]
\theoremstyle{remark}

\newcommand{\set}[1]{\left\{#1\right\}}
\newcommand{\abs}[1]{\left\vert#1\right\vert}
\newcommand{\ass}{\stackrel{\textup{\tiny def}}{=}}
\newcommand{\Prob}[2]{p( #1\rule{0pt}{12pt}\, \mid #2)}
\newcommand{\EProb}[2]{E\left[ #1\rule{0pt}{12pt}\, \mid #2\right]}
\newcommand{\EVar}[2]{\mathrm{Var}\left[ #1\rule{0pt}{12pt}\, \mid #2\right]}
\newcommand{\g}[1]{\widetilde{#1}}

\makeatletter
\patchcmd{\@settitle}{\uppercasenonmath\@title}{}{}{}
\patchcmd{\@setauthors}{\MakeUppercase}{}{}{}
\patchcmd{\section}{\scshape}{}{}{}

\makeatother

\begin{document}





\title[]{Tsirelson's bound prohibits communication through a disconnected channel}

\author[]{Avishy Carmi and Daniel Moskovich \\ \\ {\small Center for Quantum Information Science and Technology \& \\ Faculty of Engineering Sciences \\ Ben-Gurion University of the Negev, Beersheba 8410501, Israel.}}


\date{}


\begin{abstract}
Why does nature only allow nonlocal correlations up to Tsirelson's bound and not beyond? We construct a channel whose input is statistically independent of its output, but through which communication is nevertheless possible if and only if Tsirelson's bound is violated. This provides a statistical justification for Tsirelson's bound on nonlocal correlations in a bipartite setting.
\end{abstract}

\maketitle

\section{\large Introduction}

Some of the predictions made by quantum mechanics appear to be at odds with common sense. Yet quantum mechanics remains the most precisely tested and successful quantitative theory of nature. It is therefore believed that even if quantum mechanics is someday replaced, any successor will have to inherit at least some of its ``preposterous'' but highly predictive principles. Perhaps the most counter-intuitive quantum mechanical feature is \emph{nonlocality}~\citep{BellEPR}: the correlations exhibited by remote parties may exceed those allowed by any local realistic model.
%
%

The mystery of nonlocality is not only why nature is as nonlocal as it is as, but why nature is not \textit{more} nonlocal than it is. There are alternative \emph{Non-Signaling} theories which permit nonlocality beyond the quantum limit~\citep{PR1,Pop}; why doesn't nature choose one of these theories over quantum mechanics? In Section~\ref{SS:Literature} we review several previously proposed explanations. This paper presents another explanation, from statistics.

In this paper we construct a protocol (an infinite oblivious transfer) which sends messages through a disconnected channel. We show that Alice can communicate nontrivial information to Bob via this protocol if and only if the maximal quantum mechanical violation of the Bell--CHSH inequality, \emph{Tsirelson's bound}, is exceeded. We thus provide a statistical explanation of this bound that is independent of the mathematical formalism of quantum mechanics. 


We briefly recall the setting for the Bell--CHSH experiment. Section ~\ref{S:Bell} provides a more detailed account. A famous application of nonlocality is to construct an \emph{$1$-$2$ oblivious transfer protocol} between two distant agents (A)lice and (B)ob. Alice and Bob each hold a box representing one half of the quantum system to be explained.  Alice's box might, for example, contain one half of a singlet state of spin--$\frac{1}{2}$ particles, with Bob's box containing the other half~\cite{BellEPR,CHSH}. In addition, Alice possesses a pair of bits $x_0$ and $x_1$, each of which is a zero or a one. Using boolean algebra and her boxes (the protocol will be described later), Alice encodes her pair of bits into a single bit $x^{(1)}$ which she sends across a classical channel to Bob. Bob wants to know the value either of $x_0$ or of $x_1$, but Alice doesn't know which of these Bob wants to know. Bob uses the received bit $x^{(1)}$, his box, and some boolean algebra to construct an estimate $y_i$ for his desired bit $x_i$. See Figure~\ref{fig:rac} later on.

What is the probability that Bob correctly estimates the bit he wished to know? He has two possible sources of knowledge--- the  bit $x^{(1)}$ he received from Alice, and some mysterious `nonlocal' correlation between his box and Alice's. The strength of such a nonlocal coordination between two systems is captured by a parameter $c\in[-1,1]$ called the \emph{Bell--CHSH correlation}. Bob's probability of guessing the value of Alice's bit correctly is $(1+\abs{c})/2$. The \emph{Bell--CHSH inequality} states that $\abs{c}\leq 1/2$ in a world governed by classical (non-quantum) mechanics \cite{BellEPR,CHSH}. \emph{Nonlocality} is the state of affairs in which the Bell--CHSH inequality is violated. To the best of our knowledge, real world physics is nonlocal. Over the years, the violation of the Bell--CHSH inequality has been measured in increasingly accurate and loophole-free experiments, culminating in the celebrated ``loophole-free'' verification of Hensen~\textit{et.al.} \cite{Hensen:15}.


Thus, we know that $\abs{c}$ can exceed $1/2$. How large can $\abs{c}$ be? Tsirelson's bound tells us that $\abs{c}$ cannot exceed $1/\sqrt{2}$ in a world described by quantum mechanics~\cite{Tsirelson}. This quantum bound on nonlocality:
\begin{equation}\label{eq:Tsirelson}
\abs{c}\leq \frac{1}{\sqrt{2}} \enspace ,
\end{equation}
has been tested experimentally, with the current state of the art being an experiment by Kurtsiefer's~group which has achieved a value of $c$ which is only $0.0008\pm 0.00082$ distant from Tsirelson's~bound~\cite{Poh:15}. Such experimental evidence supports the contention that Tsirelson's~bound indeed holds true in the real world. Tsirelson's result as presented in the original paper is a specifically quantum mechanical fact, following from the Hilbert-space mathematical formalism for quantum mechanics, for which there has been no good conceptual physical explanation. How fundamental is Tsirelson's bound? Must this inequality also hold for any future theory which might someday supercede quantum~mechanics~\cite{Seife:05}? We are led to the following question:\\[1ex]
%
\emph{Can we identify a plausible physical principle, independent of quantum~mechanics (or independent of functional analysis), which is necessary and sufficient to guarantee that $\abs{c}\leq 1/\sqrt{2}$?}\\[1ex]

\subsection{Existing principles}\label{SS:Literature}

For the last two decades, people have searched for physical principles that bound nonlocality. It was initially expected that the physical principle of relativistic~causality (no-signaling) itself restricts the strength of nonlocality~\citep{Shimony1,Shimony2,AharonovRohrlich}. But then it was discovered that no-signaling theories may exist for which $\abs{c}>1/\sqrt{2}$. This led to the device-independent formalism of \emph{No-Signaling (NS)--boxes}~\citep{PR1,Barrett} (see also \citep{Pop}). In particular, maximum violation of the Bell--CHSH inequality is achieved by \emph{Popescu--Rohrlich (PR)--boxes} which are consistent with relativistic~causality.

So relativistic causality doesn't limit nonlocality after all; Why then does nature not permit \eqref{eq:Tsirelson} to be violated (as far as we know)? Several suggestions have been made. Superquantum correlations lead to violations of the Heisenberg uncertainty principle \citep{unc1,unc2}, which is another seemingly purely quantum result. PR--boxes would allow distributed computation to be performed with only one bit of communication \citep{vanDam}, which looks unlikely but doesn't violate any known physical law. Similarly, in stronger-than-quantum nonlocal theories some computations exceed reasonable performance limits~\citep{nonlocalcomp}. The principle of \emph{Information~Causality}~\citep{Infocause} shows that no sensible measure of mutual information exists between pairs of systems in superquantum nonlocal theories. Our approach is most directly comparable with Information~Causality, with a conceptual difference being that we use variance of an efficient estimator, therefore Fisher information, whereas information causality uses mutual information (Shannon information). The relationship between our approach and theirs is the topic of Section~\ref{sec:infocause}. Finally, it was shown that superquantum nonlocality does not permit local (non-nonlocal) physics to emerge in the limit of infinitely many microscopic systems~\citep{PRnolimit,ML}. 

\subsection{Tsirelson's bound from a statistical no-signaling condition}\label{SS:SAS}

Here we show that Tsirelson's bound follows from the following principle applied to a certain limiting Bell--CHSH setting:
%
\subsubsection*{Statistical No-Signaling}
It is impossible to communicate a nontrivial message through a channel whose output is independent of its input.\\[1ex]
%
Our strategy is to construct a channel whose input is a Bernoulli random variable $\mathbf{x}$ of mean $\theta$ and whose output is another Bernoulli random variable $\mathbf{y}$. The construction of our channel is not new--- it is a reinterpretation of the well-known van~Dam protocol~\citep{vanDam}. Through the channel, Alice sends $2^n$ samples $\mathcal{A}\ass\set{x_0,x_1,\ldots,x_{2^n-1}}$ from $\mathbf{x}$, and at the other end Bob receives a set of values $\mathcal{B}\ass\set{y_0,y_1,\ldots,y_{m-1}}$.


We imagine $\theta\in [-1,1]$ as encoding a message, perhaps in the digits of its binary expansion. Bob's task is to estimate $\theta$. The following theorem states that he can do so if and only if Tsirelson's bound fails.

\begin{thm}\label{thm:main}
\hfill \\
\begin{enumerate}
  \item The channel from $\mathbf{x}$ to $\mathbf{y}$ we construct is described by the conditional probability $p(\mathbf{y} = x \mid \mathbf{x} = x) = (1 + c^n) / 2$, where $c$ is the
 Bell--CHSH correlation. Its output satisfies $p(\mathbf{y} = 1 \mid \theta) = (1+c^n \theta) / 2$. In the $n\to \infty$ limit it disconnects for $p(\mathbf{y} \mid \mathbf{x}) = p(\mathbf{y})$ (\textit{i.e.} we can arrange that $c<1$).
  \item The unbiased estimator:
  \begin{equation*}
\hat{\theta}\ass \frac{1}{2^n c^n} \sum_{i=0}^{2^n-1} y_i \enspace ,
\end{equation*}
\noindent for $\theta$ has variance:
  \begin{equation*}
\label{eq:statnosig}
\EVar{\hat{\theta}}{\theta} = \lim_{n\to\infty}\frac{1 - c^{2n}\theta^2}{\left(2c^2\right)^n} = \left \{ \begin{array}{ll}
0, & 2c^2 > 1 \; \text{\emph{(signaling)}} \\
1, & 2c^2=1 \; \text{\emph{(randomness)}}\\
\infty, & 2c^2 < 1 \; \text{\emph{(no-signaling)}}
\end{array} \right.
\end{equation*}
  \item The estimator $\hat{\theta}$ is \emph{efficient}, \textit{i.e.} it has the minimal variance of any estimator of $\theta$ constructed from Bob's set of samples $\mathcal{B}$ for all $n\in\mathds{N}$.
\end{enumerate}
\end{thm}

\begin{figure}[htb]
\centering
\psfrag{c}[c]{}
\psfrag{a}[c]{\small $_\frac{2^n}{1-\theta^2}$}
\psfrag{b}[c]{\small $_{\frac {\left(2c^2\right)^n}{1 - c^{2n}\theta^2}}$}
\psfrag{d}[c]{\small $_{\infty}$}
\psfrag{e}[c]{\small $_{\infty \; \; \text{if} \; 2c^2 > 1}$}
\psfrag{f}[c]{\small $_{\text{$0$ or $1$}  \; \; \text{if} \; 2c^2 \leq 1}$}
\includegraphics[width=0.75\textwidth]{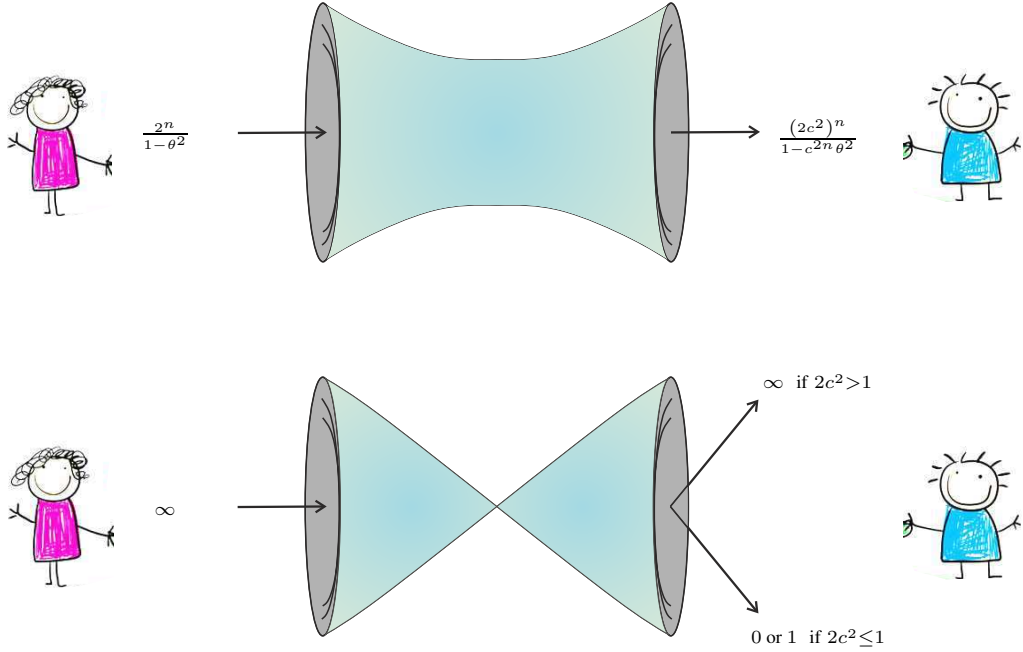}
\caption{\label{fig:rates}\small The Statistical No-Signaling condition. The van~Dam protocol defines an underlying channel which becomes disconnected in the $n\to \infty$ limit. The upper illustration shows this channel and the Fisher information (one over the variance) of the maximum likelihood estimators for $\theta$ at its input and at its output. When the number of nonlocal resources increases unboundedly, the two ends of the channel become disconnected as illustrated by a vanishing bottleneck in the lower figure. Statistical No-Signaling dictates that in this case no information can pass through. This occurs if and only if $2c^2\leq 1$. The case of $2c^2>1$ leads to a physically unreasonable limit where Bob can fully read off the value of Alice's $\theta$ through a disconnected channel.}
\end{figure}

The theorem is visually summarized by Figure~\ref{fig:rates}.

The theorem shows that failure of Tsirelson's bound leads to failure of the following consequence of Statistical No-Signaling:

\subsubsection*{Consequence of Statistical No-Signaling} In the above notation, if $\mathbf{x}$ and $\mathbf{y}$ are independent, then no estimator constructed from $\mathcal{B}$ has both mean $\theta$ and variance $0$.\\[1ex]

\subsection{Organization of this paper}
Section~\ref{S:Bell} recalls the bipartite Bell experiment and exhibits the Bell--CHSH correlation $c$ as the correlation of a certain noisy symmetric channel. Section~\ref{sec:vandam} presents the van~Dam protocol as an extension of the Bell--CHSH setup, and explain how it defines a noisy symmetric channel with correlation $c^n$. Section~\ref{sec:capacity} computes the means and variance of the estimator $\hat{\theta}$ for $\theta$, and proves that $\hat{\theta}$ is an efficient estimator. Finally, Section~\ref{sec:infocause} discusses the relationship of Statistical No-Signaling with Information Causality.

\section{\large The bipartite Bell experiment as a noisy symmetric channel}\label{S:Bell}

In this section we recall the definition of the Bell--CHSH correlation $c$ and we formulate the Bell--CHSH inequality, establishing notation. We then exhibit $c$ as the correlation of a symmetric binary channel.

\subsection{The Bell--CHSH inequality}

Let us recall the classical bipartite Bell experiment~\cite{BellEPR}. Alice and Bob each hold one half of an EPR pair (a pair of particles with certain properties summarized below) such as a singlet state of spin--$\frac{1}{2}$ particles. They each possess two different measuring instruments. Alice measures her particle using one of the instruments, and Bob measures his particles using one of his. We write $i$ for the index of the instrument used by Alice, and $a$ for its reading. Similarly, we let $j$ and $b$ denote the index of an instrument chosen by Bob and its reading correspondingly. In the language of probability, $a$ and $b$ are $\pm 1$--valued Bernoulli random variables. The choices of measuring instrument, $i$ and $j$, may be either parameters or $0/1$--valued Bernoulli random variables.

Repeating the experiment for many different EPR pairs, Alice and Bob may compute the two-point correlator $\EProb{ab}{i,j}$ of their readings $a$ and $b$ for any given pair of indices $i$ and $j$, where $E[\cdot]$ is the statistical expectation operator. We now define the \emph{Bell--CHSH correlation} $c$ by the formula:
\begin{equation}
\label{eq:c}
c \ass \frac{1}{4} \left\{ \EProb{ab}{0,0} + \EProb{ab}{0,1} +
  \EProb{ab}{1,0} - \EProb{ab}{1,1} \right\} \enspace .
\end{equation}

In a theory in which both Alice and Bob's choices, and the readings of their measuring devices, are \emph{local}, the Bell--CHSH inequality~\cite{CHSH} holds:
\begin{equation}
\label{eq:BellCHSH}
\abs{c} \leq \frac{1}{2}\enspace .
\end{equation}
Operationally speaking, locality means that Alice's readings may only be affected by her own choices (or perhaps by any other variables hidden locally at her site), and similarly for Bob's readings. Quantum mechanically, however, Alice and Bob may violate~\eqref{eq:BellCHSH}. Quantum Mechanics is thus said to be \emph{nonlocal}.

\subsection{The Bell--CHSH correlation $c$ as a channel correlation}

Non-signaling (NS)--boxes provide an abstraction and an extension of the Bell--CHSH experiment~\citep{PR1,Barrett}. This time, Alice and Bob each owns a box. Such a box may be thought of as a complete laboratory containing two measuring devices. Either participants inserts their choice of measuring device into their box. The box output is the respective reading of the chosen measuring device.

Alice and Bob share a pair of NS--boxes whose inputs are $i$ and $j$ and whose outputs are Bernoulli random variables $a$ and $b$. Take $i$, $j$, $a$, and $b$ to all be $0/1$--valued.

We will show that the Bell--CHSH correlation \eqref{eq:c} represents the correlation of a symmetric binary channel whose input is the Bernoulli random variable $\mathbf{x} \ass\, \g{ij}$ and whose output is the Bernoulli random variable $\mathbf{y} \ass\, \g{a} \cdot \g{b}$, where $\g{f}$ denotes $(-1)^f$.

Let $x \in \{0,1\}$. Define the \emph{channel correlations} $c_x$ as follows:
\begin{equation}
\label{eq:channelcorrelation}
c_{\g{x}} \ass \EProb{\mathbf{x} \mathbf{y}}{\mathbf{x}=\g{x}}= \Prob{\mathbf{y} = \g{x}}{\mathbf{x}=\g{x}} - \Prob{\mathbf{y} \neq \g{x}}{\mathbf{x}=\g{x}} = 2
\Prob{\mathbf{y}=\g{x}}{\mathbf{x}=\g{x}} - 1 \enspace.
\end{equation}
With respect to a particular choice of measuring devices $i$ and $j$, \eqref{eq:channelcorrelation} becomes:
\begin{equation}\label{eq:probs}
c_{\g{x}}(i,j) = \EProb{\g{a \oplus b} \cdot \g{ij}}{i,j, \g{ij}=\g{x}} =2 \Prob{a \oplus b = ij}{i,j,ij = x} - 1 \enspace .
\end{equation}
Pulling the condition $\g{ij} =\g{x}$ out of \eqref{eq:probs} and using $\g{a \oplus b} = \g{a} \cdot \g{b}$, we obtain:
\begin{equation}
\label{eq:corr3}
c_{\g{ij}}(i,j) =\g{ij} \cdot \EProb{\g{a} \cdot \g{b}}{i,j} = 2 \Prob{a \oplus b = ij}{i,j} - 1 \enspace .
\end{equation}

Assume the underlying channel is symmetric and therefore, $c_{\g{ij}}(i,j)$ is fixed for all $i,j$.
By \eqref{eq:corr3} the Bell--CHSH correlation \eqref{eq:c} may be written as:
\begin{equation}
\label{eq:boxprob}
c = \frac{1}{4}\left(c_1(0,0) + c_1(0,1) + c_1(1,0) + c_{-1}(1,1)\right)
= c_{\g{ij}}(i,j) = 2 \Prob{a \oplus b = ij}{i,j} - 1 \enspace.
\end{equation}
%
which is our promised interpretation of the Bell--CHSH correlation as a correlation of a noisy symmetric binary channel.

\section{\large The van~Dam protocol as a noisy symmetric channel}\label{sec:vandam}

In this section we recall the construction of the van-Dam protocol \cite{Infocause,vanDam}. We then reinterpret this protocol as underlying a noisy symmetric binary channel, as a special case of the construction of Section~\ref{S:Bell}. We compute its correlation, and establish the effect of noise on its classical component.

\subsection{The van~Dam protocol}

The van~Dam protocol realizes an \emph{oblivious transfer protocol} by means of a classical channel and a collection of NS-boxes. Each of Alice's boxes has a corresponding box on Bob's side, and different pairs of boxes are statistically independent. Suppose that Alice has in her possession the bits $x_0,\ldots,x_{m-1}$ where $m=2^n$, \; $n \geq 1$. Bob wishes to know the value of one of her bits. He may do so by specifying the address of the bit whose value he wishes to know via its binary address $j=j_{n-1} j_{n-2}\cdots j_0$. For example, if $n=2$ then Bob may specify which of the bits $x_0$ to $x_3$ he wants by specifying a binary address, $00$, $01$, $10$, or $11$. Alice bits and Bob addresses are encoded into the inputs of $2^n-1$ NS-boxes following a particular protocol which is described next.

Alice uses outputs of boxes and choices of measuring devices to determine choices of measuring devices for other boxes. Such a procedure is called \emph{wiring}. The wiring of boxes on Alice side admits a recursive description which we now give. Let $a^{k,l}_i$ denote the output of Alice's $l$th box on the $k$th level for the input $i$. Let also:
\begin{equation}
f^{k,l}\left(q_1, q_2\right) \ass q_1 \oplus a^{k,l}_{q_1 \oplus q_2}\enspace.
\end{equation}
Suppose that Alice wishes to encode $m=4$ bits with her boxes. To do so, she first picks two boxes and computes:
\begin{equation}
x^{(1)}_1 \ass f^{1,1}\left(x_0, x_1\right), \quad x^{(1)}_2 \ass f^{1,2}\left(x_2, x_3\right)\enspace .
\end{equation}
This forms the first level in her construction. The second level then follows:
\begin{equation}
x^{(2)} \ass f^{2,1}\left(x^{(1)}_1, x^{(1)}_2\right)\enspace.
\end{equation}
In this example there are only two levels and so $x^{(2)}$ is the bit which Alice transmits to Bob through the classical channel. In case where $m=2^n$ there will be $n$ levels and thus $x^{(n)}$ is the bit Bob will receive from Alice.

Unbeknownst to Alice, Bob now decides which bit $x_j$ he would like to know the value of. He takes its binary address $j=j_{n-1}j_{i-2}\cdots j_0$, and inserts $j_{k-1}$ into all of his boxes whose counterparts are on the $k$ level on Alice's side. He then uses the values $b^{k,l}_{j_{k-1}}$ that he obtains, together with the bit $x^{(n)}$ he received from Alice, to construct the decoding function:
\begin{equation}
y_j \ass x^{(n)} \oplus b^{1,l_1}_{j_0} \oplus
b^{2,l_2}_{i_1} \oplus \cdots \oplus b^{n,l_n}_{j_{n-1}}\enspace .
\end{equation}
The values $l_1,\ldots,l_n$ (which boxes Bob uses) are determined by the binary address $j=j_{n-1}j_{n-2}\cdots j_0$ via the recursive formula $l_{h-1}= 2l_h-1+l_{h-1}$ for $h=1,2,\ldots n-1$ starting from $l_n=1$.

\begin{figure}[htb]
\centering
\psfrag{s}[c]{\small $_{x_0}$}
\psfrag{t}[c]{\small $_{x_1}$}
\psfrag{m}[c]{\small $_{j}$}
\psfrag{n}[c]{\small $_{b}$}
\psfrag{a}[c]{\small $_{a}$}
\psfrag{v}[c]{\small $_{y = x^{(1)} \oplus b}$}
\psfrag{u}[l]{\small $_{x^{(1)} = x_0 \oplus a}$}
\psfrag{e}[l]{\small $_{x^{(3)} = x_0 \oplus a_6 \oplus a_4 \oplus a_0}$}
\psfrag{d}[c]{\small $_{y = x^{(3)} \oplus b_{6} \oplus b_{k} \oplus b_{l}}$}
\psfrag{1}[c]{\small $_0$}
\psfrag{2}[c]{\small $_1$}
\psfrag{3}[c]{\small $_2$}
\psfrag{4}[c]{\small $_3$}
\psfrag{5}[c]{\small $_4$}
\psfrag{6}[c]{\small $_5$}
\psfrag{7}[c]{\small $_6$}
\includegraphics[width=0.45\textwidth]{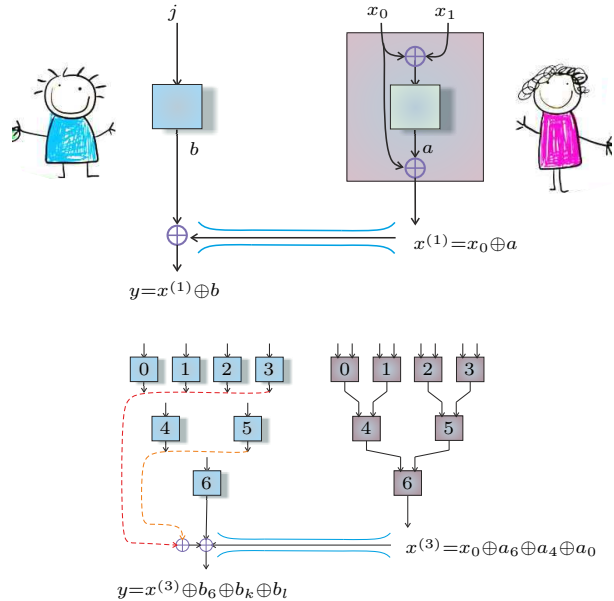}
\caption{\label{fig:rac} \small Distributed oblivious transfer (van~Dam) protocol\cite{vanDam}. Its basic building block is on the left, where Alice inserts $x_0\oplus x_1$ into her box, receives $a$, and sends $x_0\oplus a$ to Bob. Bob decides that he wants to know the value of $x_j$, and he feeds $j$ into his box, which outputs $b$. Bob's estimate of $x_i$ is then $x^{(1)}\oplus b$. When there are multiple boxes, Alice concatenates (the process is called \emph{wiring}). For example, with seven boxes, Alice begins with a collection of bits $x_0,x_1,\ldots, x_7$, and she inputs $x_{2i}\oplus x_{2i+1}$ into box $i$, where $i=0,1,2,3$, receiving $a_0, a_1, a_2, a_3$ correspondingly. The bits fed into the next level of boxes become $x^{(1)}_i\ass x_{2i}\oplus a_i$ with $i=0,1,2,3$. The final output $x^{(3)}$ is sent to Bob. Bob encodes the address of the bit he wants as the binary number $j_3j_2j_1$--- for example, if he wants $x_2$, then he sets $j_3=0$, $j_2=1$, and $j_1=0$ because $10$ is $2$ in binary. This binary encoding describes a path in his binary tree from a root to a branch, where $0$ means `go left' and $1$ means `go right'. Bob inserts $j_3$ into the lowermost box to obtain $b_{6}$. Setting $k\ass 5-(1-j_3)$, he then inserts $j_2$ into box $k$ to obtain $b_{k}$. Finally, setting $l\ass k-(3-j_3)-(1-j_2)$, Bob inserts $j_1$ into box $l$ to obtain $B_{l}$. His final estimate for $x_j$ is $y_j = x^{(3)} \oplus b_{6} \oplus b_{k} \oplus b_{l}$.}
\end{figure}

The van~Dam protocol we have described above is summarized in Figure~\ref{fig:rac}.

\subsection{The probability that Bob will guess Alice's bit correctly}

The probability that Bob will decode the correct value of the bit he desires is governed by the NS--box correlation $c$. For the simplest
case of $m=2$ where Alice and Bob share a single pair of boxes, note that
\begin{multline}
2 \Prob {y_j = x_{j_1}}{x_{j_1}}- 1 = 2 \Prob{f^{1,1}(x_0,x_1) \oplus b^{1,1}_{j_1} = x_{j_1}}{x_{j_1}} -1 \\
= 2 \Prob{x_0 \oplus a^{1,1}_{x_0 \oplus x_1} \oplus b^{1,1}_{j_1} = x_{j_1}}{x_{j_1}} - 1\enspace .
\end{multline}
As $x_{j_1}=x_0\oplus j_1(x_0\oplus x_1)$, this equals:
\begin{multline}
2 \Prob{x_0 \oplus a^{1,1}_{x_0 \oplus x_1} \oplus b^{1,1}_{j_1} = x_0 \oplus j_1
(x_0 \oplus x_1)}{x_{j_1}} - 1 \\
= 2 \Prob{a^{1,1}_{x_0 \oplus x_1} \oplus b^{1,1}_{j_1} = i j}{i=x_0
  \oplus x_1, \; j=j_1, \; x_{j_1}} - 1 = c \enspace ,
\end{multline}
which follows from \eqref{eq:boxprob}.

In general, decoding any bit out of $2^n$ possible bits involves using $n$ pairs of NS boxes. Noting that an even number of errors, $a \oplus b \neq ij$, will cancel out in such a construction, leads to~\cite{Infocause}:
\begin{equation}
\label{eq:vanDamchannel}
c^n = 2 \Prob{y_j = x_j}{x_j}- 1\enspace .
\end{equation}
We illustrate in the case that $n=2$:
\begin{multline}
\label{eq:vanDam2}
\Prob{a_{i_1}\oplus b_{j_1}\oplus a_{j_2}\oplus b_{j_2}= i_1j_1\oplus i_2j_2}{i_{1,2},j_{1,2},i_1j_1\oplus i_2j_2}=\\
\Prob{a_{i_1}\oplus b_{j_1}=i_1j_1}{a_1,b_1}\Prob{a_{i_2}\oplus b_{j_2}=i_2j_2}{i_2,j_2}+\\
\Prob{a_{i_1}\oplus b_{j_1}\neq i_1j_1}{i_1,j_1}\Prob{a_{i_2}\oplus b_{j_2}\neq i_2j_2}{i_2,j_2}= \\
\frac{1}{2}(1+c)\cdot\frac{1}{2}(1+c)+ \frac{1}{2}(1-c)\cdot \frac{1}{2}(1-c)= \frac{1}{2}(1+c^2)\enspace .
\end{multline}

\subsection{van~Dam protocol as a symmetric channel}

Assume now that instead of a string of bits, Alice has in her possession an information source that is a $\pm 1$-valued Bernoulli random variable $\mathbf{x}$
whose mean is $\theta$. Alice generates $m$ iid samples, $\tilde{x}_0,\ldots,\tilde{x}_{m-1}$ from $\mathbf{x}$ and converts them into her $0/1$-valued bits, $x_0,x_1,\ldots,x_{m-1}$ by mapping $0$ to $-1$ and $1$ to $1$. As in~\eqref{eq:vanDam2}, the van~Dam protocol has a \emph{memoriless} property:

\begin{equation}\label{eq:memoriless}
\Prob{y_i=x_i}{x_0,x_1,\ldots,x_{m-1}}= \Prob{y_i=x_i}{x_i}\enspace.
\end{equation}

From this it follows that if Alice's inputs $x_0,x_1,\ldots,x_{m-1}$ are iid then Bob's outputs $y_0,y_1,\ldots,y_{m-1}$ are also iid. Therefore the set $\tilde{y}_i\ass (-1)^{y_i}$ determine a Bernoulli random variable $\mathbf{y}$. In this way, the van~Dam protocol may be viewed as a symmetric binary channel whose input is $\mathbf{x}$ and whose output is $\mathbf{y}$. By~\eqref{eq:vanDamchannel} the channel correlation is
\begin{equation}\label{eq:vanDamcorrelation}
E\left[\mathbf{x} \mathbf{y} \mid \mathbf{x} = \tilde{x}_i \right] = 2\Prob{\mathbf{y} = \tilde{x}_i }{\mathbf{x} = \tilde{x}_i} - 1
= 2\Prob{y_i = x_i}{x_i} - 1 = c^n \enspace.
\end{equation}

\subsection{Noisy classical channel in the van~Dam protocol}

The preceding discussion of the van~Dam protocol assumed a perfect classical channel between Alice and Bob. We now relax this assumption. Let $(c^\prime)^n$ be the correlation underlying the classical channel, where $\abs{c^\prime} \leq 1$. Such a channel can be realized by concatenating $n$ copies of a noisy symmetric channel whose correlation is $c^\prime$. This correlation depends on $n$, and Alice may construct it as part of the protocol based on her knowledge of $n$.

Note first that:
\begin{multline}
\label{eq:conc}
\Prob{\mathbf{z} = i}{\mathbf{x} = i} =
\Prob{\mathbf{z} = i}{\mathbf{y} = i} \Prob{\mathbf{y} = i}{\mathbf{x} = i} +
\Prob{\mathbf{z} = i}{\mathbf{y} \neq i} \Prob{\mathbf{y} \neq i}{\mathbf{x} = i} = \\
\Prob{\mathbf{z} = i}{\mathbf{y} = i} \Prob{\mathbf{y} = i}{\mathbf{x} = i} +
\Prob{\mathbf{z} \neq i}{\mathbf{y} = i} \Prob{\mathbf{y} \neq i}{\mathbf{x} = i}\enspace.
\end{multline}
Let $\mathbf{y}$ and $\mathbf{z}$ be the input and output of a symmetric classical channel. By~\eqref{eq:channelcorrelation} we may write:
\begin{equation}\label{eq:vanDamcprimen}
(c^\prime)^n = E[\mathbf{y} \mathbf{z}] = 2\Prob{\mathbf{z} = i}{\mathbf{y} = i} - 1\enspace,
\end{equation}
and similarly we may rewrite \eqref{eq:vanDamcorrelation} as:
\begin{equation}\label{eq:vanDamn}
c^n = E\left[\mathbf{x} \mathbf{y}\right] = 2\Prob{\mathbf{y} = i}{\mathbf{x} = i} - 1\enspace.
\end{equation}
Substituting~\eqref{eq:vanDamcprimen} and~\eqref{eq:vanDamn} into~\eqref{eq:conc} gives us that for the van~Dam protocol with a noisy classical channel:
\begin{equation}\label{eq:cprimec}
\Prob{\mathbf{z} = i}{\mathbf{x} = i} = \left[ 1 + (cc^\prime)^n \right] /2 \enspace .
\end{equation}
From this we see that $(cc^\prime)^n = E\left[\mathbf{x} \mathbf{z}\right]$ is the correlation of the symmetric binary channel defined by the van~Dam protocol in the case of a classical channel with correlation $(c^\prime)^n$ and a Bell--CHSH correlation $c$.

\subsection{The van~Dam channel disconnects in the $n\to \infty $ limit}\label{sec:ChannelDisconnect}

If $\abs{c} < 1$ or $\abs{c^\prime} < 1$ then it follows that:
\begin{equation}
E[\mathbf{x} \mathbf{z}] = 2\Prob{\mathbf{z} = i}{\mathbf{x} = i} - 1 = (cc^\prime)^n \stackrel{n \to \infty}{\longrightarrow} 0\enspace.
\end{equation}
Therefore, in the $n\to \infty$ limit:
\begin{equation}\label{eq:Disconnect1}
\Prob{\mathbf{z} = i}{\mathbf{x} = i}=1/2 \enspace .
\end{equation}
But also:
\begin{equation}\label{eq:Disconnect2}
p(\mathbf{z}=i)= \Prob{\mathbf{z}=i}{\mathbf{x}=i}p(\mathbf{x}=i)+ \Prob{\mathbf{z}=i}{\mathbf{x}\neq i}p(\mathbf{x}\neq i)=
\frac{1}{2}(p(\mathbf{x}=i)+p(\mathbf{x}\neq i))= \frac{1}{2} \enspace .
\end{equation}
Combining \eqref{eq:Disconnect1} with \eqref{eq:Disconnect2} gives us that:
\begin{equation}
\Prob{\mathbf{z}}{\mathbf{x}} \stackrel{n \to \infty}{\longrightarrow}\, p(\mathbf{z}) \enspace .
\end{equation}
Thus $\mathbf{x}$ and $\mathbf{z}$ are statistically independent in the $n\to \infty$ limit, proving the first part of Theorem~\ref{thm:main}.

\section{\large Bob's estimator}\label{sec:capacity}

In Section~\ref{sec:vandam} we used the van~Dam protocol to construct a symmetric channel whose input is a $\pm 1$--valued Bernoulli random variable $\mathbf{x}$ and whose output is another $\pm 1$--valued Bernoulli random variable $\mathbf{y}$. The channel correlation is~$c^n$.

Alice sends $m$ iid random samples $\mathcal{X}\ass \set{\mathbf{x}_1,\ldots, \mathbf{x}_m}$ through the
channel. Denote the set of respective outputs $\mathcal{Y} \ass \set{\mathbf{y}_1,\ldots,
\mathbf{y}_m}$.
%
%
Assume a prior distribution for $\mathbf{x}$ given by:
\begin{equation}
\label{eq:prior}
\Prob{\mathbf{x} = -1}{\theta} = \frac{1}{2}(1 +\theta)\enspace,
\end{equation}
with parameter $\theta \in [-1,1]$.

Bob attempts to estimate $\theta$ using the estimator:
\begin{equation}\label{eq:estimator}
\hat{\theta}\ass \frac{1}{2^n c^n} \sum_{i=0}^{2^n-1} y_i \enspace .
\end{equation}
%
%
%
%
We will show that Bob's estimator is unbiased, $\EProb{\hat{\theta}}{\theta} = \theta$. Note that
\begin{equation}\label{eq:exexp}
\EProb{y_i}{\theta} = \Prob{\mathbf{y} = 1}{\theta} - \Prob{\mathbf{y} = -1}{\theta} \enspace .
\end{equation}
and
\begin{equation}\label{eq:proby}
\Prob{\mathbf{y} = -1}{\theta} = \Prob{\mathbf{y} = -1}{\mathbf{x}=-1} \Prob{\mathbf{x}=-1}{\theta} + \Prob{\mathbf{y}= -1}{\mathbf{x}=1} \Prob{\mathbf{x}=1}{\theta} =
\frac{1+c^n\theta}{2}\enspace .
\end{equation}
From~\eqref{eq:exexp} and ~\eqref{eq:proby} together, deduce:
\begin{equation}\label{eq:expy}
\EProb{y_i}{\theta} = c^n\theta \enspace .
\end{equation}
and therefore, $\EProb{\hat{\theta}}{\theta} = \theta$.

As for variance, by~\eqref{eq:expy}:
\begin{equation}\label{eq:vary}
\EVar{y_i}{\theta} = \EProb{y_i^2}{\theta} - \EProb{y_i}{\theta} ^{2} = 1 - c^{2n}\theta^2 \enspace .
\end{equation}
Therefore:
\begin{equation}\label{eq:variance}
\EVar{\hat{\theta}}{\theta} = \frac{1 - c^{2n}\theta^2}{(2c^2)^n} \enspace .
\end{equation}
We have proved the second part of Theorem~\ref{thm:main}.


\subsection{Bob's estimator $\hat{\theta}$ is efficient}\label{sec:cramerrao}

We prove efficiency of $\hat{\theta}$ by calculating the Fisher information about $\theta$ contained in Bob's set of samples $\mathcal{B}$. The Cramer--Rao Theorem tells us that one over this Fisher information is a lower bound for the variance of an estimator for $\theta$ constructed from $\mathcal{B}$. By showing that $\hat{\theta}$ saturates this bound, we will have proven that it is efficient.

We compute the Fisher information. Set $m={2^n}$. The \emph{likelihood} of $\theta$ given the set $\mathcal{B}$ is given by the expression:
\begin{equation}
\label{eq:likelihood}
\Prob{\mathcal{B}}{\theta} = \left[ \Prob{\mathbf{y} = -1}{\theta}
\right]^{\sum_{i=1}^{2^n} \mathbf{1}_{\{\mathbf{y}_i = -1\}}} \left[ \Prob{\mathbf{y} = 1}{\theta} \right]^{\sum_{i=1}^{2^n} \mathbf{1}_{\{\mathbf{y}_i = 1\}}}\enspace ,
\end{equation}
where the \emph{indicator} random variable of a random event $A$ is given as:
\begin{equation}
\mathbf{1}_A \ass \left \{
\begin{array}{ll}
1, & \text{$A$ occurred}; \\
0, & \text{otherwise}.
\end{array} \right.
\end{equation}
According to~\eqref{eq:likelihood} the log-likelihood is given by the expression:
\begin{equation}
\mathcal{L}(\theta) \ass \log \Prob{\mathcal{B}}{\theta} = \left[
  \sum_{i=1}^{2^n} \mathbf{1}_{\{\mathbf{y}_i = -1\}} \right] \log \Prob{\mathbf{y} =-1}{\theta} + \left[ \sum_{i=1}^{2^n} \mathbf{1}_{\{\mathbf{y}_i = 1\}} \right]
\log \Prob{\mathbf{y}=1}{\theta}\enspace .
\end{equation}
The \emph{Fisher information} about $\theta$ contained in the set $\mathcal{B}$ is defined as:
\begin{equation}
\label{eq:Fisher}
\mathcal{I}_{\mathcal{B}}(\theta) \ass E\left[ \left( \frac{\partial \mathcal{L}(\theta)}{\partial
      \theta} \right)^2 \right] = - E\left[ \frac{\partial^2
    \mathcal{L}(\theta) }{\partial \theta^2} \right]\enspace .
\end{equation}
Note that:
\begin{equation}
E \left[ \sum_{i=1}^{2^n} \mathbf{1}_{\{\mathbf{y}_i = s\}} \right] = \sum_{i=1}^{2^n}
E \left[ \mathbf{1}_{\{\mathbf{y}_i = s\}} \right] = {2^n} p(\mathbf{y} = s \mid \theta),
\quad s=-1,1 \enspace .
\end{equation}
Using this, \eqref{eq:Fisher} reads:
\begin{equation}
\label{eq:Fishersym}
\mathcal{I}_{\mathcal{B}}(\theta) = \frac{(2c^2)^n}{1 - c^{2n} \theta^2} \enspace.
\end{equation}
%
%
Indeed the Fisher information about $\theta$ in $\mathcal{B}$ as given by Equation~\eqref{eq:Fishersym} equals one over the variance of $\hat{\theta}$ as given by Equation~\eqref{eq:variance}. Thus, by the Cramer--Rao Theorem, $\hat{\theta}$ is an efficient estimator for $\theta$. Parenthetically, note that the minimum of $\mathcal{I}_{\mathcal{B}}(\theta)$ is obtained for
$\theta=0$ in which case $p(\mathbf{x} \mid \theta)=1/2$ and $\mathcal{I}_{\mathcal{B}}(0)=(2c^2)^n$. We have proved the final part of Theorem~\ref{thm:main}.

\section{\large Relation to Information Causality}
\label{sec:infocause}


Of previous non-quantum justifications of Tsirelson's bound, Information Causality (IC) is perhaps the closest to Statistical No-Signalling~\cite{Infocause}. IC is also stated as a limit on communication:\\[1ex]
\emph{
Information gain that Bob can reach about a previously unknown to him data set of Alice, by using all his local resources and $m$ classical bits communicated by Alice, is at most $m$ bits.
}\\[1ex]
IC is formally a restriction on the classical channel capacity. Detecting violation of this principle therefore requires the prominence of nonlocal resources, which the authors achieve through the application of IC to the van~Dam protocol, that is the same communication protocol used in this paper.

Formally, the authors define an Information Causality quantity $I$ as the Shannon mutual information of Alice's input and Bob's output given the value of the single bit transmitted in the van~Dam protocol. IC holds if $I\leq 1$ and is violated if $I>1$. At the end of the supplementary section of, the authors prove the expression:
\begin{equation}
\label{eq:IC}
I \geq \frac{1}{2 \ln (2)} \left(c_1^2 + c_{-1}^2\right)^n \enspace ,
\end{equation}
where $c_i \ass \EProb{\mathbf{x} \mathbf{y}}{\mathbf{x}=\tilde{i}}$ as in \eqref{eq:channelcorrelation}. In the symmetric setting, $c_1 = c_{-1} = c$, and for $\theta = 0$, \eqref{eq:IC} and \eqref{eq:Fishersym} combine to yield:
 \begin{equation}
\label{eq:IC2}
I \geq \frac{2^nc^{2n}}{2 \ln (2)} = \frac{[1 - c^{2n} \theta^2] \mathcal{I}_{\mathcal{B}}(\theta)}{2 \ln (2)} \enspace .
\end{equation}
In particular, in the $N\to \infty$ limit, if $2c^2>1$ then $\mathcal{I}_{\mathcal{B}}(\theta) \to \infty$ implying that $I\to \infty$. Thus, violation of Statistical No-Signaling implies violation of IC. Conversely, as \eqref{eq:IC} is an inequality, it is unknown whether Tsirelson's bound being satisfied implies $I\leq 1$ (IC for the van~Dam protocol), although, by our main theorem, it does imply $\mathcal{I}_{\mathcal{B}}(\theta)\leq 1$ (Statistical No-Signaling for the van~Dam protocol).


\section{\large Conclusions}
\label{sec:discussion}

We have formulated a \emph{Statistical No-Signaling} principle which dictates that no information can pass through a disconnected channel.
A violation of Tsirelson's bound, \textit{i.e.} a value of $\abs{c}$ greater that $1/\sqrt{2}$, allows us to violate Statistical No-Signalling by constructing an asymptotically disconnected channel through which Bob can construct an unbiased estimator with variance $0$ for Alice's parameter $\theta$. Conversely, when Tsirelson's bound holds, then, through this channel, so does Statistical No-Signalling. Our construction thus provides a purely statistical justification for Tsirelson's bound, independent of quantum mechanics.

\section*{}
\begin{description}
\small
\item[Acknowledgements.] The authors thank D. Rohrlich for useful discussions.



\end{description}


%

\end{document}